\documentclass[submission,copyright,creativecommons]{eptcs} 
\providecommand{\event}{ICE 2024}

\RequirePackage{underscore} 
\RequirePackage[T1]{fontenc} 

\usepackage{stmaryrd}
\RequirePackage{amsthm}
\RequirePackage{thmtools}
\declaretheorem{theorem}
\declaretheorem{definition}

\declaretheorem{example}

\input{macros.tex}

\PassOptionsToPackage{capitalise,noabbrev}{cleveref}
\RequirePackage{cleveref}

\title{A Gentle Overview of Asynchronous Session-based Concurrency: Deadlock Freedom by Typing}
\author{%
    Bas van den Heuvel
    \institute{Karlsruhe University of Applied Sciences, Karlsruhe, and University of Freiburg, Freiburg, Germany}
    \and
    Jorge A.\ P\'erez
    \institute{University of Groningen, The Netherlands}
}
\def\titlerunning{Asynchronous Session-Based Concurrency}
\def\authorrunning{B.\ van den Heuvel, J.A.\ P\'erez}

\begin{document}

\maketitle

\begin{abstract}
While formal models of concurrency tend to focus on synchronous communication, asynchronous communication is relevant in practice. In this paper, we will discuss asynchronous communication in the context of session-based concurrency, the model of computation in which \emph{session types} specify the structure of the two-party protocols implemented by the channels of a communicating process.
We overview recent work on addressing the challenge of ensuring the \emph{deadlock-freedom} property for message-passing processes that communicate asynchronously in cyclic process networks governed by session types.
We offer a  gradual presentation of three typed process frameworks and outline how they may be used to guarantee deadlock freedom for a concurrent functional language with sessions.
\end{abstract}


\section{Introduction}
\label{s:intro}

The purpose of this paper is to overview recent work on new verification techniques that enforce deadlock freedom for \emph{message-passing processes}. We target concurrent systems that form cyclic process networks and that communicate \emph{asynchronously}, governed by protocols expressed as \emph{session types}~\cite{conf/parle/TakeuchiHK94,conf/esop/HondaVK98}.

We rely on \emph{process calculi} as formal models of concurrency, as they provide a firm foundation for specifying and analyzing message-passing programs and for principled designs of programming abstractions involving concurrent, interactive behavior. We are also interested in asynchronous communication, which, from the standpoint of distributed systems, can be informally described as the kind of process communication in which no global clock is assumed; as such, an observer has no way of knowing if the message they have sent has been received.
These intuitions can be precisely formulated in a language-independent way (cf.\ \cite{conf/concur/Selinger97}).
Unsurprisingly, asynchronous communication is of clear practical relevance: it is the standard in most distributed systems and web-based applications nowadays.

Despite this pragmatic interest, process calculi such as the $\pi$-calculus tend to focus on synchronous communication, rather than on asynchronous communication. In fact, as we discuss later on, the study of formalisms such as the asynchronous $\pi$-calculus originated from an interest in the essential ingredients of the synchronous $\pi$-calculus. There is a vast literature on the asynchronous $\pi$-calculus as a `fundamental core' of the $\pi$-calculus, its behavioral theory, and its relationship with the original synchronous $\pi$-calculus, in particular from the point of view of relative (or comparative)  expressiveness (see, e.g.,~\cite{conf/ecoop/HondaT91,report/Boudol92,journal/mscs/Palamidessi03,journal/tcs/AmadioCS98,journal/iandc/NestmannP00,journal/iandc/Nestmann00,conf/lics/PalamidessiSVV06,journal/tcs/CacciagranoCP07}). In the $\pi$-calculus, asynchronous communication admits an elegant and economical formulation:  asynchronous processes can be defined simply by decreeing that in output-prefixed processes $\fProc{\pOut x[v]; P}$ the continuation $\fProc{P}$ can only correspond to the inactive process $\fProc{0}$. This ensures that there are no processes that are blocked by an output action, but also that a process $\fProc{\pOut x[v];0}$ can be regarded as an `output particle' that has been emitted but not yet received by some intended receiver. These particles could then be arranged into some suitable structure, such as a queue or a stack~\cite{chapter/BeauxisPV08}.

Asynchronous communication is also relevant for \emph{session-based concurrency}, which can be described as the model of interaction in which processes exchange messages following some predetermined protocols specified as session types. Session types specify sequences of input and output actions, possibly recursive, which define \emph{communication structures} between two or more interacting parties.
Session-type systems leverage those structures to ensure  that interacting processes always respect their intended specifications and never exhibit issues such as message mismatches (e.g., ill-formatted data), message duplication and loss, out-of-order messages, race conditions, and \emph{deadlocks}---the insidious situation in which processes are permanently blocked, awaiting indefinitely a message that will never arrive. The use of session types for excluding deadlocks in asynchronous processes 
is the central theme in our work.

Session-based concurrency has been widely studied using (variants of) the $\pi$-calculus, which provide a simple yet rigorous framework for developing verification techniques for message-passing programs.
In this context, asynchronous communication usually has a  less economical definition than in the untyped setting:
processes/programs are typically defined together with some runtime entities, such as buffers, which explicitly account for the in-transit messages by following the structure of their corresponding sessions~\cite{journal/iandc/Dezani-CiancagliniDMY09,journal/jfp/GayV10}. Different designs for these buffers, accounting for different levels of granularity, are possible~\cite{conf/forte/KouzapasYH11}. This treatment has direct consequences on the notions of causality that govern reasoning over well-typed processes: {actions from different sessions should be independent from each other, but actions {(including outputs)} within a session should follow the ordering described by their session type}.
As a result, the machinery required by asynchronous sessions   entails some notational burden, for instance when formulating the meta-theoretical results for well-typed processes. Moving to an asynchronous setting has also important consequences for central notions, such as \emph{subtyping}, which is decidable under a synchronous semantics but becomes undecidable in the asynchronous case~\cite{conf/fossacs/LangeY17}.

In this paper, we are interested in verification techniques for ensuring that asynchronous session processes are \emph{deadlock free}. Just as session-based concurrency integrates elements and concepts originating from different areas (concurrency theory, process calculi, type systems, programming languages), we consider an amalgamation of two separate developments, which leads to a clean formulation of asynchronous communication in which deadlock-freedom guarantees hold for a wide class of processes.

On the one hand, we consider logical correspondences, in the style of Curry-Howard, that connect session types and linear logic~\cite{conf/concur/CairesP10,conf/icfp/Wadler12}. This line of work provides in particular clean foundations for both the analysis of deadlock freedom and for asynchronous communication. Indeed, as shown by DeYoung \etal~\cite{conf/csl/DeYoungCPT12}, in a logically-motivated setting, asynchronous communication enjoys a remarkably economical  formulation, in which output particles represent the intended communication structure using continuation passing. On the other hand, we consider type systems for the $\pi$-calculus that exclude deadlocks by considering \emph{priority-based} approaches~\cite{conf/concur/Kobayashi06,conf/lics/Padovani14}, which avoid vicious cycles in advanced communication patterns. These two strands of work are distinct in nature but complementary nevertheless; in particular, it is known that priority-based approaches to deadlock freedom are strictly more powerful than logic-based approaches~\cite{conf/express/DardhaP15,journal/jlamp/DardhaP22}. In this paper we show how the proposed amalgamation enables a fresh understanding of the key insights involved in enforcing deadlock freedom in an asynchronous setting, gradually going from no enforcement, to enforcement restricted to tree-like topologies, and culminating in deadlock freedom for the cyclic topologies of processes that abound in practice (such as those present in parallel algorithms).

\paragraph{Structure of the document.}
\Cref{s:async} gives a high-level discussion on the interplay of asynchronous communication and (session) protocols.
\Cref{s:ap} presents \AP: a $\pi$-calculus with asynchronous communication whose session-type system enforces conformance to session protocols but does not exclude deadlocks.
\Cref{s:acp,s:apcp} build upon \AP to gradually illustrate the essentials of deadlock freedom by typing. We first introduce \ACP, an asynchronous variant of Wadler's \CP~\cite{conf/icfp/Wadler12} that enforces deadlock freedom for processes that form tree-like networks. Then, we present the key ideas underlying \APCP, an {enhancement of \AP} with a priority-based approach to typing, which enforces deadlock freedom also for processes that form cyclic networks.
\Cref{s:lastn} {briefly discusses} \LASTn, a functional language with asynchronous sessions (based on the language by Gay and Vasconcelos~\cite{journal/jfp/GayV10}), for which deadlock freedom can be guaranteed via a correct translation into \APCP.
\Cref{s:concl} collects some final remarks.

\paragraph{Origin of the results.}
This paper is intended as a gentle introduction to our journal paper~\cite{journal/lmcs/vdHeuvelP24}, which offers a full treatment of \APCP, its meta-theoretical results, formal connections with \LASTn, and comparisons with related works.
In particular,
\Cref{s:apcp,s:lastn} collect selected results first reported in~\cite{journal/lmcs/vdHeuvelP24}.
For the sake of presentation, here we consider typed calculi without recursive processes and recursive session types, which are included in~\cite{journal/lmcs/vdHeuvelP24}.   The discussion in \Cref{s:async} and the asynchronous variant of \CP presented in \Cref{s:acp} are new to this presentation.

\section{Encodings as (Session) Protocols and Asynchronous Communication}
\label{s:async}

In the theory of the $\pi$-calculus, asynchronous communication was not the first choice. The $\pi$-calculus was introduced as a calculus of
synchronous, channel-based communication, from which asynchrony arose as a (syntactic) limitation---a sort of afterthought.
Much of what we know about asynchronous communication in this setting actually comes from  studies investigating the (non-)existence  of (correct) \emph{encodings} of synchronous into asynchronous communication. Perhaps unsurprisingly, the theory of session types followed a similar path: it was first formulated using programming models with synchronous communication; the interest in asynchrony came later, and continues to be relevant, especially as session types have found their way into (mainstream) programming languages~\cite{journal/ftpl/AnconaBB0CDGGGH16}.

{Encodings} between process calculi can intuitively  be seen as \emph{protocols}: given a step in a source calculus, an encoding gives a precise sequence of steps in the target calculus that represents it.
A bit of history may be instructive here.
Shortly after the (synchronous) $\pi$-calculus was introduced,  researchers sought to determine the \emph{essential ingredients} of interaction and concurrency. An initial subject of study was \emph{polyadicity}---the ability to send and receive finite lists of names in a single communication step:
$$\fProc{\pOut x[v_1, \ldots, v_k]; P \| \pIn x(y_1, \ldots, y_k); Q} \redd \fProc{P \| Q\pSub{ v_1/y_1 } \ldots \pSub{ v_k/y_k }} $$
where $\fProc{\pOut x[\tilde{v}]}$ and $\fProc{\pIn x(\tilde{y})}$ denote output and input prefixes, respectively, and `$\fProc{;}$' and `$\fProc{\|}$' denote sequential and parallel composition, respectively.
The question is then whether the polyadic $\pi$-calculus can be encoded into the \emph{monadic} variant, in which at most one value can be exchanged. Milner~\cite{chapter/Milner93} gave the following protocol for the exchange of a list of values $\fProc{v_1}, \ldots, \fProc{v_k}$ over name $\fProc{x}$: create a fresh name $\fProc{s}$, send~$\fProc{s}$ over~$\fProc{x}$, and then use $\fProc{s}$ to individually transmit each $\fProc{v_i}$ (with $\fProc{i} \in 1 \ldots k$) using a monadic communication. Hence, this protocol represents a single $k$-adic communication with $k+1$ monadic communications, using $\fProc{s}$ as a private session to avoid interferences. Interestingly, as simple and plausible as this protocol looks, its correctness is not obvious, in particular if one considers \emph{full abstraction}: as shown by Quaglia and Walker~\cite{journal/iandc/QuagliaW05}, using types for monadic processes is essential for establishing a sound and complete correspondence between source (polyadic) processes and their corresponding target (monadic) processes.

This brings us back to the issue of asynchronous communication.
Studies on asynchronous variants of the $\pi$-calculus originated from the question: can a synchronous communication discipline be represented in the simpler and more pragmatic asynchronous setting, in which output is not a blocking operation? For the $\pi$-calculus without choice constructs, two encodings/protocols were independently proposed by Boudol~\cite{report/Boudol92} and by Honda and Tokoro~\cite{conf/ecoop/HondaT91} (who studied it for a core language for objects).

We briefly review these two encodings, following the presentation by Van Glabbeek~\cite{journal/ipl/Glabbeek18}, writing $\fProc{\Boudol{\cdot}}$ and $\fProc{\HondaT{\cdot}}$, respectively (in both cases, the encoding is a homomorphism for other process constructs):
\begin{align*}
    \fProc{\Boudol{\pOut x[y];P}}  &= \fProc{\pRes{u}(\pOut x[u] \| \pIn u(w);(\pOut w[y] \| \Boudol{P}))}
    & \fProc{\HondaT{\pOut x[y];P}}  &= \fProc{\pIn x(u);(\pOut u[y] \| \HondaT{P})}
	\\
    \fProc{\Boudol{\pIn x(z);P}} &= \fProc{\pIn x(u);\pRes{w}(\pOut u[w] \| \pIn w(z); \Boudol{P})}
    & \fProc{\HondaT{\pIn x(z);P}} &= \fProc{\pRes{w}(\pOut x[w] \| \pIn w(z); \HondaT{P})}
\end{align*}
The encodings adopt different approaches to represent synchronous communication. In Boudol's encoding/protocol, a single communication step is represented using three steps in the asynchronous calculus: first on $\fProc{x}$ and then on the two fresh names $\fProc{u}$ and $\fProc{w}$ (on which the source value $\fProc{y}$ is finally communicated). Observe that the direction of actions is preserved: an output is encoded into an output, and same for input. Honda and Tokoro's encoding/protocol is simpler.  It only involves one fresh name but it does not preserve directionality: indeed, the encoding of input takes the initiative by sending a freshly created name on which the communication of $\fProc{y}$ will occur. Despite these differences, both encodings are correct with respect to Gorla's correctness criteria~\cite{journal/iandc/Gorla10}, as shown in~\cite{journal/ipl/Glabbeek18}.

In our opinion, the view of encodings as protocols is insightful in itself, but also because we are interested in session types as a way of specifying protocols for (deadlock-free) concurrent processes. In this respect, there are two salient points worth making.
\begin{itemize}
	\item
First, there is a tension between asynchronous communication and session types: while the former aims at \emph{unconstraining} behaviors (by ensuring that only inputs are blocking points in communication), session types aim at \emph{constraining} behaviors, for a good reason: to ensure that processes conform to some intended communication structure. This tension does not entail a conflict, but it does have a consequence: the incorporation of session types in asynchronous process calculi results in a model that stands ``in between'' synchrony and asynchrony but differs from both~\cite{conf/forte/KouzapasYH11}: (output) actions from different sessions should be independent from each other, but (output) actions within a session should follow the ordering described by their session type.
\item Second, it is known that the choice between synchronous and asynchronous communication directly influences deadlock-freedom analyses~\cite{conf/fmoods/CoppoDY07}: the asynchronous setting appears as the most convenient scenario in which to develop techniques for ensuring deadlock freedom for (session) processes, as it involves the least amount of blocking operations. Also, when considering sessions, asynchronous communication makes differences between different sessions even more prominent. To see this, consider the following process in a synchronous setting:
$$ \fProc{P} = \fProc{\pRes{xy}\pRes{uw}( \pOut u[z] ; \pOut y[v]; P_1 \|  \pIn x(v); \pIn w(z) ; P_2)}$$
where we use the restriction $\fProc{\pRes{xy}}$ to declare $\fProc{x}$ and $\fProc{y}$ as dual endpoints of a session.
Hence, in $\fProc{P}$ we have that $\fProc{x}$ and $\fProc{y}$ form one session, different from the session formed by $\fProc{u}$ and $\fProc{w}$. Also, $P$ consists of two sub-processes, both with blocking operations (outputs in the left sub-process, inputs in the right sub-process); the cyclic dependencies induced by these operations make $P$ deadlocked.

Having an output on a session that blocks an output from another session is difficult to justify. While a variant of $\fProc{P}$ in which these two outputs are swapped (as in $\fProc{\pOut y[v]; \pOut u[z] ;  P_1}$) would immediately solve the issue, a more fundamental observation is that $\fProc{P}$ is not directly expressible in an asynchronous setting. In fact, in an asynchronous setting we could have a process like
$$ \fProc{\pRes{xy}\pRes{uw}( \pOut u[z]  \| \pOut y[v] \| P_1  \| \pIn x(v); \pIn w(z) ; P_2  )}$$
that expresses the dependency between the two sessions in a deadlock-free manner.

\end{itemize}
\noindent
Having presented a high-level discussion on the interplay between encodings/protocols, asynchronous communication, and deadlock freedom, we now move on to formally presenting a basic process model of asynchronous communication and its corresponding type system.

\section{Asynchronous Processes}
\label{s:ap}

\begin{figure}[t!]{The \AP process language: syntax and reduction semantics.}{f:acp:proc}
    Process syntax:
    \begin{align*}
        \fProc{P}, \fProc{Q} &
        \begin{array}[t]{@{}l@{}lr@{\kern2em}l@{\kern1em}lr@{}}
            {} ::= {} &
            \fProc{\pOut x[a,b]} & \text{send}
            & \sepr &
            \fProc{\pIn x(y,z) ; P} & \text{receive}
            \\ \sepr* &
            \fProc{\pSel x[b] < j} & \text{selection}
            & \sepr &
            \fProc{\pBra x(z) > {\{ i : P \}_{i \in I}}} & \text{branch}
            \\ \sepr* &
            \fProc{P \| Q} & \text{parallel}
            & \sepr &
            \fProc{\pRes{xy} P} & \text{restriction}
            \\ \sepr* &
            \fProc{0} & \text{inaction}
            & \sepr &
            \fProc{\pFwd [x<>y]} & \text{forwarder}
        \end{array}
    \end{align*}

    \dashes

    Structural congruence:
    \begin{mathpar}
        \begin{bussproof}[sc-alpha]
            \bussAssume{
                \fProc{P} \alpheq \fProc{Q}
            }
            \bussUn{
                \fProc{P} \sc \fProc{Q}
            }
        \end{bussproof}
        \and
        \begin{bussproof}[sc-par-unit]
            \bussAx{
                \fProc{P \| 0} \sc \fProc{P}
            }
        \end{bussproof}
        \and
        \begin{bussproof}[sc-par-comm]
            \bussAx{
                \fProc{P \| Q} \sc \fProc{Q \| P}
            }
        \end{bussproof}
        \and
        \begin{bussproof}[sc-par-assoc]
            \bussAx{
                \fProc{P \| ( Q \| R )} \sc \fProc{( P \| Q ) \| R}
            }
        \end{bussproof}
        \and
        \begin{bussproof}[sc-scope]
            \bussAssume{
                \fProc{x},\fProc{y} \notin \fn(\fProc{P})
            }
            \bussUn{
                \fProc{P \| \pRes{xy} Q} \sc \fProc{\pRes{xy} ( P \| Q )}
            }
        \end{bussproof}
        \and
        \begin{bussproof}[sc-res-comm]
            \bussAx{
                \fProc{\pRes{xy} \pRes{zw} P} \sc \fProc{\pRes{zw} \pRes{xy} P}
            }
        \end{bussproof}
        \and
        \begin{bussproof}[sc-res-symm]
            \bussAx{
                \fProc{\pRes{xy} P} \sc \fProc{\pRes{yx} P}
            }
        \end{bussproof}
        \and
        \begin{bussproof}[sc-res-inact]
            \bussAx{
                \fProc{\pRes{xy} 0} \sc \fProc{0}
            }
        \end{bussproof}
        \and
        \begin{bussproof}[sc-fwd-symm]
            \bussAx{
                \fProc{\pFwd [x<>y]} \sc \fProc{\pFwd [y<>x]}
            }
        \end{bussproof}
        \and
        \begin{bussproof}[sc-res-fwd]
            \bussAx{
                \fProc{\pRes{xy} \pFwd [x<>y]} \sc \fProc{0}
            }
        \end{bussproof}
    \end{mathpar}

    \dashes

    Reduction:
    \begin{mathpar}
        \begin{bussproof}[red-send-recv]
            \bussAx{
                \fProc{\pRes{xy} ( \pOut x[a,b] \| \pIn y(a',b') ; Q )}
                \redd
                \fProc{Q \pSub{a/a',b/b'}}
            }
        \end{bussproof}
        \and
        \begin{bussproof}[red-sel-bra]
            \bussAssume{
                \fProc{j} \in \fProc{I}
            }
            \bussUn{
                \fProc{\pRes{xy} ( \pSel x[b] < j \| \pBra y(b') > {\{ i : Q_i \}_{i \in I}} )}
                \redd
                \fProc{Q_j \pSub{b/b'}}
            }
        \end{bussproof}
        \and
        \begin{bussproof}[red-fwd]
            \bussAssume{
                \fProc{y} \neq \fProc{z}
            }
            \bussUn{
                \fProc{\pRes{xy} ( \pFwd [x<>z] \| P )}
                \redd
                \fProc{P \pSub{z/y}}
            }
        \end{bussproof}
        \and
        \begin{bussproof}[red-sc]
            \bussAssume{
                \fProc{P} \sc \fProc{P'}
            }
            \bussAssume{
                \fProc{P'} \redd \fProc{P'}
            }
            \bussAssume{
                \fProc{Q'} \equiv \fProc{Q}
            }
            \bussTern{
                \fProc{P} \redd \fProc{Q}
            }
        \end{bussproof}
        \and
        \begin{bussproof}[red-res]
            \bussAssume{
                \fProc{P} \redd \fProc{Q}
            }
            \bussUn{
                \fProc{\pRes{xy} P} \redd \fProc{\pRes{xy} Q}
            }
        \end{bussproof}
        \and
        \begin{bussproof}[red-par]
            \bussAssume{
                \fProc{P} \redd \fProc{Q}
            }
            \bussUn{
                \fProc{P \| R} \redd \fProc{Q \| R}
            }
        \end{bussproof}
    \end{mathpar}
\end{figure}

We start by presenting a calculus of \emph{asynchronous processes}, dubbed \AP. We define its syntax, reduction semantics, and session-type system. Well-typed processes perform their ascribed session protocols but may run into deadlocks. As such, \AP provides a basic framework for developing more sophisticated typing disciplines that enforce deadlock freedom, namely \ACP and \APCP, to be introduced later on.

\paragraph{Syntax.}
We write $\fProc{a},\fProc{b},\fProc{c},\ldots,\fProc{x},\fProc{y},\fProc{z},\ldots$ to denote \emph{names} (or \emph{endpoints}); by convention we use the early letters of the alphabet for the objects of output-like constructs.
Also, we write $\fProc{\tilde{x}},\fProc{\tilde{y}},\fProc{\tilde{z}},\ldots$ to denote finite sequences of names.
With a slight abuse of notation, we sometimes write $\fProc{x_i} \in \fProc{\tilde{x}}$ to refer to a specific element in the sequence~$\fProc{\tilde{x}}$.
Also, we write $\fProc{i},\fProc{j},\fProc{k},\ldots$ to denote \emph{labels} for choices, and $\fProc{I},\fProc{J},\fProc{K},\ldots$ to denote finite sets of labels.
In \AP, communication is
\emph{asynchronous} (cf.~\cite{conf/ecoop/HondaT91,conf/occ/HondaT91,report/Boudol92}) and
\emph{dyadic}: each communication involves the transmission of a pair of names, usually interpreted as a message name and a continuation name.
We use $\fProc{P},\fProc{Q},\ldots$ to denote processes.

\Cref{f:acp:proc} (top) gives the syntax of processes.
The send $\fProc{\pOut x[a,b]}$ emits along $\fProc{x}$ a message name $\fProc{a}$ and a continuation name $\fProc{b}$.
The receive~$\fProc{\pIn x(y,z) ; P}$ blocks until along $\fProc{x}$ a message and continuation name are received (referred to in~$\fProc{P}$ as the placeholders $\fProc{y}$ and $\fProc{z}$, respectively), binding $\fProc{y}$ and $\fProc{z}$ in $\fProc{P}$.
The selection~\mbox{$\fProc{\pSel x[b] < i}$} sends along $\fProc{x}$ a label $\fProc{i}$ and a continuation name $\fProc{b}$.
The branch $\fProc{\pBra x(z) > {\{ i : P_i \}_{i \in I}}}$ blocks until it receives along $\fProc{x}$ a label $\fProc{i} \in \fProc{I}$ and a continuation name (referred to in $\fProc{P_i}$ as the placeholder~$\fProc{z}$), binding $\fProc{z}$ in each $\fProc{P_i}$.
In the rest of this paper, we refer to sends, receives, selections, and branches as \emph{prefixes} (even though sends and selections do not prefix a continuation process).
We refer to sends and selections collectively as \emph{outputs}, and to receives and branches as \emph{inputs}.

The process $\fProc{P \| Q}$ denotes the parallel composition of $\fProc{P}$ and $\fProc{Q}$.
Restriction $\fProc{\pRes{xy} P}$ binds $\fProc{x}$ and $\fProc{y}$ in $\fProc{P}$, thus declaring them as the two names of a channel and enabling communication (cf.\ \cite{journal/iandc/Vasconcelos12}).
The process~$\fProc{0}$ denotes inaction.
The forwarder $\fProc{\pFwd [x<>y]}$ is a primitive copycat process that links together names $\fProc{x}$ and~$\fProc{y}$.

Names
are free unless otherwise stated (i.e., unless they are bound somehow).
We write $\fn(\fProc{P})$
for the set of free names
of $\fProc{P}$,
and $\bn(\fProc{P})$ for the set of bound names of $\fProc{P}$.
Also, we write $\fProc{P \pSub{ x/y }}$ to denote the capture-avoiding substitution of the free occurrences of $\fProc{y}$ in $\fProc{P}$ for $\fProc{x}$.
We write sequences of substitutions $\fProc{P \pSub{ x_1/y_1 } \ldots \pSub{ x_n/y_n }}$ as $\fProc{P \pSub{ x_1/y_1, \ldots, x_n/y_n }}$.




\paragraph{Reduction semantics.}
The reduction relation for processes ($\fProc{P} \redd \fProc{Q}$) formalizes how complementary outputs/inputs on connected names may synchronize.
As usual for \picalc -calculi, reduction relies on \emph{structural congruence} ($\fProc{P} \sc \fProc{Q}$), which relates processes with minor syntactic differences.
Structural congruence is the smallest congruence on the syntax of processes (\Cref{f:acp:proc}~(top)) satisfying the axioms in \Cref{f:acp:proc}~(middle).

Structural congruence defines the following properties for processes.
Processes are equivalent up to $\alpha$-equivalence (Rule~\ruleLabel{sc-alpha}).
Parallel composition is associative (Rule~\ruleLabel{sc-par-assoc}) and commutative (Rule~\ruleLabel{sc-par-comm}), with unit $\fProc{0}$ (Rule~\ruleLabel{sc-par-unit}).
A parallel process may be moved into or out of a restriction as long as the bound channels do not occur free in the moved process (Rule~\ruleLabel{sc-scope}): this is \emph{scope inclusion} and \emph{scope extrusion}, respectively.
Restrictions on inactive processes may be dropped (Rule~\ruleLabel{sc-res-inact}), and the order of names in restrictions and of consecutive restrictions does not matter (Rules~\ruleLabel{sc-res-symm} and~\ruleLabel{sc-res-comm}, respectively).
Forwarders are symmetric (Rule~\ruleLabel{sc-fwd-symm}), and equivalent to inaction if both names are bound together through restriction (Rule~\ruleLabel{sc-res-fwd}).

We define the reduction relation $\fProc{P} \redd \fProc{Q}$ by the axioms and closure rules in \Cref{f:acp:proc} (bottom).
We write $\redd*$ for the reflexive, transitive closure of $\redd$.
Rule~$\ruleLabel{red-send-recv}$ synchronizes a send and a receive on connected names and substitutes the message and continuation names.
Rule~$\ruleLabel{red-sel-bra}$ synchronizes a selection and a branch:
the received label determines the continuation process, substituting the continuation name appropriately.
Rule~$\ruleLabel{red-fwd}$ implements the forwarder as a substitution.
Rules~$\ruleLabel{red-sc}$, $\ruleLabel{red-res}$, and~$\ruleLabel{red-par}$ close reduction under structural congruence, restriction, and parallel composition, respectively.

\paragraph{Type system.}
\AP types processes by assigning binary session types to names.
Following Curry-Howard interpretations, we present session types as linear logic propositions (cf., e.g., Caires \etal~\cite{journal/mscs/CairesPT16}, Wadler~\cite{conf/icfp/Wadler12}, Caires and P\'erez~\cite{conf/esop/CairesP17}, and Dardha and Gay~\cite{conf/fossacs/DardhaG18}).

\begin{definition}[Session Types for \AP]
    \label{d:acp:types}
    The following grammar defines the syntax of \emph{session types} $\fType{A},\fType{B}$.
    \begin{align*}
        \fType{A},\fType{B}
        &::= \fType{A \tensor B} \sepr \fType{A \parr B} \sepr \fType{\oplus \{ i : A \}_{i \in I}} \sepr \fType{\with \{ i : A \}_{i \in I}} \sepr \fType{\bullet}
    \end{align*}
\end{definition}

\noindent
A name of type $\fType{A \tensor B}$ (resp.\ $\fType{A \parr B}$) first sends (resp.\ receives) a message name of type $\fType{A}$ and a continuation name of type~$\fType{B}$.
A name of type $\fType{\oplus \{ i : A_i \}_{i \in I}}$ selects a label $\fType{i} \in \fType{I}$ and sends a continuation name of type~$\fType{A_i}$.
A name of type $\fType{\with \{ i : A_i \}_{i \in I}}$ offers a choice: after receiving a label $\fType{i} \in \fType{I}$, the continuation name should behave as $\fType{A_i}$.
We write $\fType{\bullet}$ to denote the type of a session protocol that is finished, i.e., a session that is \emph{closed}.
Closed sessions are usually typed using the linear logic units $\fType{1}$ and $\fType{\bot}$ (cf., e.g., \cite{conf/concur/CairesP10,conf/fossacs/DardhaG18}), but \AP does not associate any process behavior with closed sessions, so we follow Caires~\cite{report/Caires14} in conflating $\fType{1}$ and~$\fType{\bot}$ to the single, self-dual type $\fType{\bullet}$.

\emph{Duality}, the cornerstone notion of session types and linear logic, ensures that the two names of a channel have complementary behaviors.

\begin{definition}[Duality]
    \label{d:acp:duality}
    The \emph{dual} of session type $\fType{A}$, denoted $\fType{\ol{A}}$, is defined inductively as follows:
    \begin{align*}
        \fType{\ol{A \tensor B}}
        &\deq
        \fType{\ol{A} \parr \ol{B}}
        &
        \fType{\ol{\oplus \{ i : A_i \}_{i \in I}}}
        &\deq
        \fType{\with \{ i : \ol{A_i} \}_{i \in I}}
        &
        \fType{\ol{\bullet}}
        &\deq
        \fType{\bullet}
        \\
        \fType{\ol{A \parr B}}
        &\deq
        \fType{\ol{A} \tensor \ol{B}}
        &
        \fType{\ol{\with \{ i : A_i \}_{i \in I}}}
        &\deq
        \fType{\oplus \{ i : \ol{A_i} \}_{i \in I}}
    \end{align*}
\end{definition}

Judgments are of the form $$\fType{\fProc{P} \tAP \Gamma}$$ where
$\fProc{P}$ is a process
and
$\fType{\Gamma}$ is a context that records assignments of types to channels of the form $\fType{\fProc{x}:A}$.
A judgment $\fType{\fProc{P} \tAP \Gamma}$ then means that $\fProc{P}$ can be typed in accordance with the type assignments for names recorded in $\fType{\Gamma}$.
The context~$\fType{\Gamma}$ obeys \emph{exchange} (assignments may be silently reordered),
but disallows \emph{weakening} (all assignments must be used, except names typed with~$\fType{\bullet}$) and \emph{contraction} (assignments may not be duplicated).
The empty context is written $\fType{\tEmpty}$.
In writing $\fType{\Gamma, \fProc{x}:A}$ we assume that $\fProc{x} \notin \dom(\fType{\Gamma})$.

\begin{figure}[t!]{The typing rules of \AP.}{f:acp:typing}
    \begin{mathpar}
        \begin{bussproof}[typ-send]
            \bussAx{
                \fType{\fProc{\pOut x[a,b]} \tAP \fProc{x}:A \tensor B, \fProc{a}:\ol{A}, \fProc{b}:\ol{B}}
            }
        \end{bussproof}
        \and
        \begin{bussproof}[typ-recv]
            \bussAssume{
                \fType{\fProc{P} \tAP \Gamma, \fProc{y}:A, \fProc{z}:B}
            }
            \bussUn{
                \fType{\fProc{\pIn x(y, z) ; P} \tAP \Gamma, \fProc{x}:A \parr B}
            }
        \end{bussproof}
        \and
        \begin{bussproof}[typ-sel]
            \bussAssume{
                \fType{j} \in \fType{I}
            }
            \bussUn{
                \fType{\fProc{\pSel x[b] < j} \tAP \fProc{x}:\oplus \{ i : A_i \}_{i \in I}, \fProc{b}:\ol{A_j}}
            }
        \end{bussproof}
        \and
        \begin{bussproof}[typ-bra]
            \bussAssume{
                \forall \fType{i} \in \fType{I} \colon \fType{\fProc{P_i} \tAP \Gamma, \fProc{z}:A_i}
            }
            \bussUn{
                \fType{\fProc{\pBra x(z) > {\{ i : P_i \}_{i \in I}}} \tAP \Gamma, \fProc{x}:\with \{ i : A_i \}_{i \in I}}
            }
        \end{bussproof}
        \and
        \begin{bussproof}[typ-end]
            \bussAssume{
                \fType{\fProc{P} \tAP \Gamma}
            }
            \bussUn{
                \fType{\fProc{P} \tAP \Gamma, \fProc{x}:\bullet}
            }
        \end{bussproof}
        \and
        \begin{bussproof}[typ-par]
            \bussAssume{
                \fType{\fProc{P} \tAP \Gamma}
            }
            \bussAssume{
                \fType{\fProc{Q} \tAP \Delta}
            }
            \bussBin{
                \fType{\fProc{P \| Q} \tAP \Gamma, \Delta}
            }
        \end{bussproof}
        \and
        \begin{bussproof}[typ-res]
            \bussAssume{
                \fType{\fProc{P} \tAP \Gamma, \fProc{x}:A, \fProc{y}:\ol{A}}
            }
            \bussUn{
                \fType{\fProc{\pRes{xy} P} \tAP \Gamma}
            }
        \end{bussproof}
        \and
        \begin{bussproof}[typ-inact]
            \bussAx{
                \fType{\fProc{0} \tAP \tEmpty}
            }
        \end{bussproof}
        \and
        \begin{bussproof}[typ-fwd]
            \bussAx{
                \fType{\fProc{\pFwd [x<>y]} \tAP \fProc{x}:\ol{A}, \fProc{y}:A}
            }
        \end{bussproof}
    \end{mathpar}
\end{figure}

\Cref{f:acp:typing} gives the typing rules.
We describe the typing rules from a \emph{bottom-up} perspective.
Rule~\ruleLabel{typ-send} types a send; this rule does not have premises to provide a continuation process, leaving the free message and continuation names to be bound to a continuation process using Rules~\ruleLabel{typ-par} and~\ruleLabel{typ-res} (both of which will be discussed next).
Similarly, Rule~\ruleLabel{typ-sel} types a selection, where the continuation name is free.
Rules~\ruleLabel{typ-recv} and~\ruleLabel{typ-bra} type receives and branches, respectively.

In the tradition of simply-typed $\pi$-calculi~\cite{book/SangiorgiW01}, we have two separate rules for parallel composition and restriction.
Rule~\ruleLabel{typ-par} types the parallel composition of two processes that do not share assignments on the same names.
Rule~\ruleLabel{typ-res} types a restriction, where the two restricted names must be of dual type.
Rule~\ruleLabel{typ-end} implements a constrained form of weakening, which silently removes a closed name from the typing context.
Rule~\ruleLabel{typ-inact} types an inactive process with no names.
Rule~\ruleLabel{typ-fwd} types forwarding between names of dual type.

\AP satisfies an important form of type soundness that guarantees consistency of typing across structural congruence and reduction.
This property is key to proving safety properties such as session fidelity (correct implementation of assigned session types) and communication safety (no message mismatches).

\begin{theorem}[Type Preservation for \AP]
    Given $\fType{\fProc{P} \tAP \Gamma}$ and $\fProc{Q}$ such that $\fProc{P} \sc \fProc{Q}$ or $\fProc{P} \redd \fProc{Q}$, we have $\fType{\fProc{Q} \tAP \Gamma}$.
\end{theorem}

Deadlock freedom is a fundamental property for message-passing processes.
However, typing in \AP is too permissive to guarantee deadlock freedom, as illustrated by the following example.

\begin{example}
    \label{x:ap:deadlock}
    Consider the process $$\fProc{\pRes{xy} \pRes {uw} ( \pIn x(v,x') ; \pOut u[a,b] \| \pIn w(z,w') ; \pOut y[c,d] )}$$
    which can be considered as the ``hello world'' of deadlocked message-passing processes.
    The process is deadlocked because the left receive (on $\fProc{x}$) is waiting for the right send (on $\fProc{y}$), which is blocked by a receive (on $\fProc{w}$) waiting for the left send (on $\fProc{u}$), which is blocked by the left receive.
    We refer to such a state as a \emph{cyclic dependency}.
    The  corresponding typing derivation, given below, is valid in \AP; for brevity, the right subtree (analogous to the left one) is omitted and superscripts on rule labels indicate a number of repeated applications of the same rule.
    \[
        \begin{bussproof}
            \bussAx{
                \fType{\fProc{\pOut u[a,b]} \tAP \fProc{u}:\bullet \tensor \bullet , \fProc{a}:\bullet , \fProc{b}:\bullet}
            }
            \bussUn[\ruleLabel{typ-end}[2]]{
                \fType{\fProc{\pOut u[a,b]} \tAP \fProc{v}:\bullet , \fProc{x'}:\bullet , \fProc{u}:\bullet \tensor \bullet , \fProc{a}:\bullet , \fProc{b}:\bullet}
            }
            \bussUn[\ruleLabel{typ-recv}]{
                \fType{\fProc{\pIn x(v,x') ; \pOut u[a,b]} \tAP \begin{array}[t]{@{}l@{}}
                    \fProc{x}:\bullet \parr \bullet , \fProc{u}:\bullet \tensor \bullet , \\
                    \fProc{a}:\bullet , \fProc{b}:\bullet
                \end{array}}
            }
            \bussAssume{
                \vdots
            }
            \noLine
            \bussUn{
                \fType{\fProc{\pIn w(z,w') ; \pOut y[c,d]} \tAP \begin{array}[t]{@{}l@{}}
                    \fProc{w}:\bullet \parr \bullet , \fProc{y}:\bullet \tensor \bullet , \\
                    \fProc{c}:\bullet , \fProc{d}:\bullet
                \end{array}}
            }
            \bussBin[\ruleLabel{typ-par}]{
                \fType{\fProc{\pIn x(v,x') ; \pOut u[a,b] \| \pIn w(z,w') ; \pOut y[c,d]} \tAP \fProc{x}:\bullet \parr \bullet , \fProc{u}:\bullet \tensor \bullet , \fProc{a}:\bullet , \fProc{b}:\bullet , \fProc{w}:\bullet \parr \bullet , \fProc{y}:\bullet \tensor \bullet , \fProc{c}:\bullet , \fProc{d}:\bullet}
            }
            \bussUn[\ruleLabel{typ-res}[2]]{
                \fType{\fProc{\pRes{xy} \pRes {uw} ( \pIn x(v,x') ; \pOut u[a,b] \| \pIn w(z,w') ; \pOut y[c,d] )} \tAP \fProc{a}:\bullet , \fProc{b}:\bullet , \fProc{c}:\bullet , \fProc{d}:\bullet}
            }
        \end{bussproof}
    \]
    Note that deadlock freedom is usually guaranteed for \emph{closed} processes, i.e., without free names.
    The process above is not closed, but it can be trivially closed because all its free names are typed $\fType{\bullet}$.
\end{example}

In fact, the cyclic dependencies introduced in the example above are the only source of deadlock in our process calculus.
As we will see in the next two sections, there are multiple ways for typing to guarantee deadlock freedom by ruling out cyclic dependencies.

\section{Asynchronous \CP}
\label{s:acp}

\ACP is an asynchronous variant of Wadler's Classical Processes (\CP)~\cite{conf/icfp/Wadler12}.
It can be obtained from \AP by a minor yet crucial modification to \Cref{f:acp:typing}: Rules~\ruleLabel{typ-par} and~\ruleLabel{typ-res} are replaced by the following Rule~\ruleLabel{typ-cut}, which combines parallel composition and restriction:
\[
    \begin{bussproof}
        \bussAssume{
            \fType{\fProc{P} \tAP \Gamma , \fProc{x}:A}
        }
        \bussAssume{
            \fType{\fProc{Q} \tAP \Delta , \fProc{y}:\ol{A}}
        }
        \bussBin[\ruleLabel{typ-cut}]{
            \fType{\fProc{\pRes{xy} ( P \| Q )} \tAP \Gamma , \Delta}
        }
    \end{bussproof}
\]
As we will see, replacing Rules~\ruleLabel{typ-par} and~\ruleLabel{typ-res} with Rule~\ruleLabel{typ-cut} guarantees deadlock freedom simply by ruling out all possible cyclic dependencies, because it ensures that every pair of processes shares at most a single pair of names with dual behaviors.
We shall write $\fType{\tACP}$ instead of $\fType{\tAP}$ to denote the difference between the two type systems.

The different typing for parallel composition has an effect on the semantics of (typable) processes. Indeed, note that most structural congruence rules in \cref{f:acp:proc} (middle) do not preserve typing under~$\fType{\tACP}$ (e.g., the left process in Rule~\ruleLabel{sc-scope} is not typable at all).
Therefore, we  define an alternative structural congruence for \ACP, denoted $\scCut$, with Rules~\ruleLabel{sc-alpha} and~\ruleLabel{sc-fwd-symm} as in \cref{f:acp:proc}~(middle) and the following rules:
\begin{mathpar}
    \begin{bussproof}[sc-cut-symm]
        \bussAx{
            \fProc{\pRes{xy} ( P \| Q )} \scCut \fProc{\pRes{yx} ( Q \| P )}
        }
    \end{bussproof}
    \and
    \begin{bussproof}[sc-cut-assoc-L]
        \bussAssume{
            \fProc{y} \notin \fn(\fProc{Q})
        }
        \bussAssume{
            \fProc{w} \notin \fn(\fProc{P})
        }
        \bussBin{
            \fProc{\pRes{xy} ( P \| \pRes{zw} ( Q \| R ) )}
            \scCut
            \fProc{\pRes{zw} ( Q \| \pRes{xy} ( P \| R ) )}
        }
    \end{bussproof}
    \and
    \begin{bussproof}[sc-cut-assoc-R]
        \bussAssume{
            \fProc{y} \notin \fn(\fProc{R})
        }
        \bussAssume{
            \fProc{z} \notin \fn(\fProc{P})
        }
        \bussBin{
            \fProc{\pRes{xy} ( P \| \pRes{zw} ( Q \| R) )}
            \scCut
            \fProc{\pRes{zw} ( \pRes{xy} ( P \| Q ) \| R )}
        }
    \end{bussproof}
\end{mathpar}
Accordingly, we define reduction for \ACP, denoted $\reddCut$, as in \cref{f:acp:proc}~(bottom), except that Rule~\ruleLabel{red-sc} uses $\scCut$.
We write $\fProc{P} \nreddCut$ whenever there exists no $\fProc{Q}$ such that $\fProc{P} \reddCut \fProc{Q}$.
We have the following:

\begin{theorem}[Type Preservation for \ACP]
    Given $\fType{\fProc{P} \tACP \Gamma}$ and $\fProc{Q}$ such that $\fProc{P} \scCut \fProc{Q}$ or $\fProc{P} \reddCut \fProc{Q}$, we have $\fType{\fProc{Q} \tACP \Gamma}$.
\end{theorem}

\begin{example}
    \label{x:acp:deadlockFree}
    Recall the deadlocked process from \cref{x:ap:deadlock}.
    Under $\fType{\tACP}$, this process is not typable, because its two parallel sub-processes are connected on two pairs of names.
    We can create a well-typed variant by splitting one of the sub-processes into a parallel composition instead of sequence, such that the resulting two sub-processes can be connected to the original left sub-process on a single pair of names each:
    $$
    \fProc{\pRes{xy} \big( \pRes{uw} ( \pIn x(v,x') ; \pOut u[a,b] \| \pIn w(z,w') ; 0 ) \| \pOut y[c,d] \big)}
    $$
    The corresponding derivation is shown below; for readability, we omit the left subtree, which is identical to that in \cref{x:ap:deadlock} (writing $\fType{\tACP}$ instead of $\fType{\tAP}$).
    \[
        \begin{bussproof}
            \def\defaultHypSeparation{\hskip3px}
            \bussAssume{
                \vdots
            }
            \noLine
            \bussUn{
                \fType{\fProc{\pIn x(v,x') ; \pOut u[a,b]} \tACP \begin{array}[t]{@{}l@{}}
                    \fProc{x}:\bullet \parr \bullet , \fProc{u}:\bullet \tensor \bullet , \\
                    \fProc{a}:\bullet , \fProc{b}:\bullet
                \end{array}}
            }
            \bussAx[\ruleLabel{typ-inact}]{
                \fType{\fProc{0} \tACP \tEmpty}
            }
            \bussUn[\ruleLabel{typ-end}[2]]{
                \fType{\fProc{0} \tACP \fProc{z}:\bullet , \fProc{w'}:\bullet}
            }
            \bussUn[\ruleLabel{typ-recv}]{
                \begin{array}[t]{@{}l@{}}
                    \fType{\fProc{\pIn w(z,w') ; 0} \tACP \fProc{w}:\bullet \parr \bullet}
                    \\
                \end{array}
            }
            \bussBin[\ruleLabel{typ-cut}]{
                \begin{array}[t]{@{}l@{}}
                    \fType{\fProc{\pRes{uw} ( \pIn x(v,x') ; \pOut u[a,b] \| \pIn w(z,w') ; 0 ) } \tACP \fProc{x}:\bullet \parr \bullet , \fProc{a}:\bullet , \fProc{b}:\bullet}
                    \\
                \end{array}
            }
            \bussAx[\ruleLabel{typ-send}]{
                \fType{\fProc{\pOut y[c,d]} \tACP \begin{array}[t]{@{}l@{}}
                    \fProc{y}:\bullet \tensor \bullet , \\
                    \fProc{c}:\bullet , \fProc{d}:\bullet
                \end{array}}
            }
            \bussBin[\ruleLabel{typ-cut}]{
                \fType{\fProc{\pRes{xy} \big( \pRes{uw} ( \pIn x(v,x') ; \pOut u[a,b] \| \pIn w(z,w') ; 0 ) \| \pOut y[c,d] \big)} \tACP \fProc{a}:\bullet , \fProc{b}:\bullet , \fProc{c}:\bullet , \fProc{d}:\bullet}
            }
        \end{bussproof}
    \]
    As a result, the cyclic dependency from \cref{x:ap:deadlock} is broken, and the process is deadlock free.
    Note that we use $\fProc{0}$ to accommodate the standalone receive.
\end{example}

The following result captures common definitions of deadlock freedom (cf., e.g., \cite{conf/concur/Kobayashi06}), where, e.g., when a process contains a non-blocked output it will reduce until an input on an opposite endpoint is non-blocked and communication can take place.

\begin{theorem}[Deadlock Freedom for \ACP]
    \label{t:acp:df}
    Given $\fType{\fProc{P} \tACP \tEmpty}$, if $\fProc{P} \nreddCut$, then $\fProc{P} \scCut \fProc{0}$.
\end{theorem}

\paragraph{Commuting conversions.}

    The sequent calculus of linear logic induces a form of proof equivalence known as \emph{commuting conversions}, where rule applications may be commuted past each other while preserving assumptions and conclusion.
    Caires and Pfenning~\cite{conf/concur/CairesP10} noticed that some commuting conversions correspond to structural congruences in type systems such as \ACP.
    For example, Rule~\ruleLabel{sc-cut-assoc-L} commutes two applications of Rule~\ruleLabel{typ-cut}.
    On the other hand, other commuting conversions induce more significant process transformations, e.g., they change the order of blocking prefixes.
    Writing $\comm$ to denote such transformations and annotating processes with relevant free names, we have, e.g.,
    \[
        \fProc{\pRes{xy} ( \pRes{aa'} \pRes{bb'} \pOut z[a,b] ; ( P_{a'} \| Q_{b',x} ) \| R_{y} )}
        \comm
        \fProc{ \pRes{aa'} \pRes{bb'} \pOut z[a,b] ; ( P_{a'} \| \pRes{xy} ( Q_{b',x} \| R_{y} ) ) }
    \]
    by commuting the application of the synchronous variant of Rule~\ruleLabel{typ-send} past the application of Rule~\ruleLabel{typ-cut}.
    These conversions do not correspond to structural congruences but to typed behavioral equivalences (cf.~\cite{conf/esop/PerezCPT12,journal/iandc/PerezCPT14}).
        Importantly, this behavioral characterization of commuting conversions  holds under synchronous communication, where outputs are blocking.
    Interestingly, DeYoung \etal~\cite{conf/csl/DeYoungCPT12} discovered that, under asynchronous communication, some of these latter commuting conversions that involve outputs correspond to simple structural congruences. This is the case for our previous example:
    \[
        \fProc{\pRes{xy} ( \pRes{bb'} ( \pRes{aa'} ( \pOut z[a,b] \| P_{a'} ) \| Q_{b',x} ) \| R_{y} )}
        \scCut
        \fProc{ \pRes{bb'} ( \pRes{aa'} ( \pOut z[a,b] \| P_{a'} ) \| \pRes{xy} ( Q_{b',x} \| R_{y} ) ) }
    \]
As discussed in \cref{s:async}, to study deadlock freedom at its core, it is desirable to have a setting with the least possible ``amount of blockage''.
Hence, the above discoveries confirm that deadlock freedom is best studied under asynchronous communication, for synchronous communication entails unnecessary blockages by outputs leading to artificial sources of deadlock, whereas asynchronous communication adequately considers inputs as the only source of blockage.

\section{Priorities for \AP}
\label{s:apcp}
We now consider \APCP~\cite{journal/lmcs/vdHeuvelP24}, which  uses the same process language as \AP: syntax, structural congruence, and reduction are defined exactly as in \cref{f:acp:proc}.
Following~\cite{conf/concur/Kobayashi06,conf/fossacs/DardhaG18},
we extend the session types of \AP (\cref{d:acp:types}) with \emph{priority} annotations on binary connectives.
Priorities are natural numbers. Intuitively, prefixes whose type has lower priority should not be blocked by those with higher priority.

We write $\fPri{\pi},\fPri{\rho},\ldots$ to denote priorities, and $\fPri{\omega}$ to denote the ultimate priority that is greater than all other priorities  and cannot be increased further.
That is, for every $\fPri{\pi} \in \mathbb{N}$, $\fPri{\omega > \pi}$ and $\fPri{\omega + \pi} = \fPri{\omega}$.
Also, by abuse of notation, we reuse $\fType{A},\fType{B},\ldots$ to denote \APCP session types.

\begin{definition}[Session Types for \APCP]
    \label{d:apcp:types}
    The following grammar defines the syntax of \emph{session types}~$\fType{A},\fType{B}$.
    \begin{align*}
        \fType{A},\fType{B}
        &::= \fType{A \tensor^{\fPri{\pi}} B} \sepr \fType{A \parr^{\fPri{\pi}} B} \sepr \fType{\oplus^{\fPri{\pi}} \{ i : A \}_{i \in I}} \sepr \fType{\with^{\fPri{\pi}} \{ i : A \}_{i \in I}} \sepr \fType{\bullet}
    \end{align*}
\end{definition}

Session types retain the same meaning as in \cref{s:acp}.
A name of type $\fType{\bullet}$ does not require a priority, as closed names do not exhibit behavior and thus are non-blocking.

The priority of a type is determined by the priority of its outermost connective:

\begin{definition}[Priorities]
    \label{d:apcp:priority}
    For session type $\fType{A}$, $\fPri{\pr(\fType{A})}$ denotes its \emph{priority}:
    \begin{align*}
        \fPri{\pr(\fType{A \tensor^{\fPri{\pi}} B})}
        \deq
        \fPri{\pr(\fType{A \parr^{\fPri{\pi}} B})}
        \deq
        \fPri{\pr(\fType{\oplus^{\fPri{\pi}} \{ i : A_i \}_{i \in I}})}
        \deq
        \fPri{\pr(\fType{\with^{\fPri{\pi}} \{ i : A_i \}_{i \in I}})}
        &\deq
        \fPri{\pi}
        &
        \fPri{\pr(\fType{\bullet})}
        \deq
        \fPri{\omega}
    \end{align*}
\end{definition}

The priority of $\fType{\bullet}$ is the constant $\fPri{\omega}$: the type denotes the ``final'', non-blocking part of protocols.
Although the connectives $\fType{\tensor}$ and $\fType{\oplus}$ also denote non-blocking prefixes, they do block their continuation until they are received.
Hence, their priority is not constant.

\APCP typing judgments are denoted $\fType{\fProc{P} \tAPCP \Gamma}$.
We write $\fPri{\pr(\fType{\Gamma})}$ to denote the least of the priorities of all types in $\fType{\Gamma}$:
\begin{align*}
    \fPri{\pr(\fType{\tEmpty})} &\deq \fPri{\omega}
    &
    \fPri{\pr(\fType{\Gamma , \fProc{x}:A})} &\deq \fPri{\bmin(\pr(\fType{\Gamma}),\pr(\fType{A}))}.
\end{align*}

\begin{figure}[t!]{Typing rules of \APCP that add priority conditions to the \AP typing rules in \cref{f:acp:typing}.}{f:apcp:typing}
    \begin{mathpar}
        \begin{bussproof}[typ-send]
            \bussAssume{
                \fPri{\pi < \pr(\fType{A}),\pr(\fType{B})}
            }
            \bussUn{
                \fType{\fProc{\pOut x[y,z]} \tAPCP \fProc{x}:A \tensor^{\fPri{\pi}} B, \fProc{y}:\ol{A}, z:\ol{B}}
            }
        \end{bussproof}
        \and
        \begin{bussproof}[typ-recv]
            \bussAssume{
                \fType{\fProc{P} \tAPCP \Gamma, \fProc{y}:A, \fProc{z}:B}
            }
            \bussAssume{
                \fPri{\pi < \pr(\fType{\Gamma})}
            }
            \bussBin{
                \fType{\fProc{\pIn x(y, z) ; P} \tAPCP \Gamma, \fProc{x}:A \parr^{\fPri{\pi}} B}
            }
        \end{bussproof}
        \and
        \begin{bussproof}[typ-sel]
            \bussAssume{
                \fType{j} \in \fType{I}
            }
            \bussAssume{
                \fPri{\pi < \pr(\fType{A_j})}
            }
            \bussBin{
                \fType{\fProc{\pSel x[z] < j} \tAPCP \fProc{x}:\oplus^{\fPri{\pi}} \{ i : A_i \}_{i \in I}, \fProc{z}:\ol{A_j}}
            }
        \end{bussproof}
        \and
        \begin{bussproof}[typ-bra]
            \bussAssume{
                \forall \fType{i} \in \fType{I} \colon \fType{\fProc{P_i} \tAPCP \Gamma, \fProc{z}:A_i}
            }
            \bussAssume{
                \fPri{\pi < \pr(\fType{\Gamma})}
            }
            \bussBin{
                \fType{\fProc{\pBra x(z) > {\{ i : P_i \}_{i \in I}}} \tAPCP \Gamma, \fProc{x}:\with^{\fPri{\pi}} \{ i : A_i \}_{i \in I}}
            }
        \end{bussproof}
    \end{mathpar}
\end{figure}

The typing rules of \APCP ensure that prefixes with lower priority are not blocked by those with higher priority.
To this end, they enforce the following laws:
\begin{enumerate}

    \item\label{i:law:sendSel}
        Outputs with priority $\fPri{\pi}$ must have messages and continuations with priority strictly larger than $\fPri{\pi}$;

    \item\label{i:law:recvBra}
        A prefix typed with priority $\fPri{\pi}$ must be prefixed only by inputs with priority strictly smaller than $\fPri{\pi}$;

    \item\label{i:law:resFwd}
        Dual prefixes leading to a synchronization must have equal priorities.

\end{enumerate}
The typing rules for \APCP are the same as those for \AP in \cref{f:acp:typing}, and have the same meaning.
To enforce the laws above, conditions on priorities are needed: \cref{f:apcp:typing} shows the modified typing rules for prefixes.
Rules~\ruleLabel{typ-send} and~\ruleLabel{typ-sel} require that the priority of the subject is lower than the priorities of both objects (continuation and payload)---this enforces Law~\labelcref{i:law:sendSel}.
In Rules~\ruleLabel{typ-recv} and~\ruleLabel{typ-bra}, the used name's priority must be lower than the priorities of the other types in the continuation's typing context---this enforces Law~\labelcref{i:law:recvBra}.
Law~\labelcref{i:law:resFwd} is enforced by extending type duality (\cref{d:acp:duality}) to require that $\fType{A = \ol{B}}$ not only if the sequences of actions in $\fType{A}$ and $\fType{B}$ are complementary, but also that their priority annotations match up perfectly.
This is then implicitly enforced by Rules~\ruleLabel{typ-res} and~\ruleLabel{typ-fwd} (omitted from \cref{f:apcp:typing}).

We have the following results:
\begin{theorem}[Type Preservation for \APCP]
    Given $\fType{\fProc{P} \tAPCP \Gamma}$ and $\fProc{Q}$ such that $\fProc{P} \sc \fProc{Q}$ or $\fProc{P} \redd \fProc{Q}$, we have $\fType{\fProc{Q} \tAPCP \Gamma}$.
\end{theorem}

\begin{theorem}[Deadlock Freedom for \APCP]
    \label{t:apcp:df}
    Given $\fType{\fProc{P} \tAPCP \tEmpty}$, if $\fProc{P} \nredd$, then $\fProc{P} \sc \fProc{0}$.
\end{theorem}

\begin{example}
    Recall the deadlocked process from \cref{x:ap:deadlock}:
    \[
        \fProc{\pRes{xy} \pRes{uw} ( \pIn x(v, x') ; \pOut u[a,b] \| \pIn w(z, w') ; \pOut y[c,d] )}
    \]
    An attempt to type this process under $\fType{\tAPCP}$ reveals the cyclic dependency: the dual receive on $\fProc{x}$ and send on $\fProc{y}$ have priority $\fPri{\pi}$, the dual send on $\fProc{u}$ and receive on $\fProc{w}$ have priority $\fPri{\rho}$, and the two applications of Rule~\ruleLabel{typ-recv} contradictorily require $\fPri{\pi < \rho}$ and $\fPri{\rho < \pi}$.

    On the other hand, recall the non-deadlocked variant from \cref{x:acp:deadlockFree}:
    \[
        \fProc{\pRes{xy} \big( \pRes{uw} ( \pIn x(v,x') ; \pOut u[a,b] \| \pIn w(z,w') ; 0 ) \| \pOut y[c,d] \big)}
    \]
    In this case, $\fType{\tAPCP}$ only requires $\fPri{\pi < \rho}$ but not $\fPri{\rho < \pi}$, so no cyclic dependency is detected and the process is considered well typed.
\end{example}

By abuse of notation, we write $\fType{\tAP}$, $\fType{\tACP}$, and $\fType{\tAPCP}$ to denote the \emph{sets of processes} that are well typed by the typing rules of \AP, \ACP, and \APCP, respectively. These classes are related by strict inclusions:

\begin{theorem}[Comparative Expressiveness]
    We have $\fType{\tACP} \subset \fType{\tAPCP} \subset \fType{\tAP}$.
\end{theorem}

The example above is typable under $\fType{\tACP}$, so we illustrate how much more expressive \APCP is than \ACP with a variant of Milner's cyclic scheduler~\cite{book/Milner89} (inspired by~\cite[Example~1]{conf/fossacs/DardhaG18}), which is not typable under $\fType{\tACP}$.
This variant is finite and has a fixed number of three participants; the interested reader can consult~\cite{journal/lmcs/vdHeuvelP24} for a recursive cyclic scheduler with $n \geq 2$ participants.

\begin{example}
    We construct a ring of three partial schedulers $\fProc{A_i}$, each of which simultaneously invokes a worker $\fProc{B_i}$ and waits for the others to finish.
    Each scheduler $\fProc{A_i}$ ($1 \leq i \leq 3$) communicates with its worker on name $\fProc{w_i}$ (connected to the worker's name $\fProc{z_i})$, with its left neighbor on name $\fProc{x_{i-1}}$ ($\fProc{x_0} \deq \fProc{x_3}$), and with its right neighbor on name $\fProc{y_i}$.
    The schedulers and workers are defined as follows, writing~`$\fProc{\_}$' for unused names:
    \begin{align*}
        \fProc{A_1} &\deq \fProc{\begin{array}[t]{@{}l@{}}
            \pRes{a_1 y'_1} \pRes{b_1 w'_1} \pRes{c_1 \_}
            \\
            \big(
                \pSel y_1[a_1] < start \| \pSel w_1[b_1] < start \| \pBra w'_1(\_) > \{ done : \pSel y'_1[c_1] < done \| \pBra x_3(x'_3) > \{ start : \pBra x'_3(\_) > \{ done : 0 \} \mkern-2mu\} \mkern-2mu\}
            \mkern-2mu\big)
        \end{array}}
        \\
        \fProc{A_i} &\deq \fProc{\begin{array}[t]{@{}l@{}}
            \pRes{a_i y'_i} \pRes{b_i w'_i} \pRes{c_i \_}
            \\
            \big(
                \pBra x_{i-1}(x'_{i-1}) > \{ start : \pSel w_i[b_i] < start \| \pSel y_i[a_i] < start \| \pBra w'_i(\_) > \{ done : \pBra x'_{i-1}(\_) > \{ done : \pSel y'_i[c_i] < done \} \} \}
            \big)
        \end{array}}
        \tag*{$2 \leq i < 3$}
        \\
        \fProc{B_i} &\deq \fProc{\pRes{d_i \_} ( \pBra z_i(z'_i) > \{ start : doSomething() ; \pSel z'_i[d_i] < done \} )}
        \tag*{$1 \leq i \leq 3$}
    \end{align*}
    Thus, $\fProc{A_1}$ starts the routine by signaling its right neighbor and worker.
    It then waits for its worker to finish, and signals its right neighbor it is done.
    Only in the end does it wait for a start signal from its left neighbor, and then wait for it to finish.
    The other two schedulers are defined identically.
    They wait for a start signal from their left neighbor, and then start their worker.
    After its worker is done, it signals to its right neighbor, and waits for its left neighbor to finish.
    Workers wait for a start signal, do some task ($\fProc{doSomething()}$ stands for some arbitrary computation), and signal that they are finished afterwards.

    A complete, cyclic scheduler $\fProc{Sched}$ is then formed by connecting the partial schedulers on channels between $\fProc{x_i}$ and $\fProc{y_i}$ for $1 \leq i \leq 3$.
    \[
        \fProc{Sched} \deq \fProc{\pRes{x_1 y_1} \pRes{x_2 y_2} \pRes{x_3 y_3} \big( \pRes{w_1 z_1} ( A_1 \| B_1 ) \| \pRes{w_2 z_2} ( A_2 \| B_2 ) \| \pRes{w_3 z_3} ( A_3 \| B_3 ) \big)}
    \]

    To type $\fProc{Sched}$, we assign the following types to names for $1 \leq i \leq 3$:
    \begin{itemize}
        \item $\fType{\fProc{w_i}:\oplus^{\fPri{\pi_i}} \{ start : \with^{\fPri{\pi'_i}} \{ done : \bullet \} \}}$,
        \item $\fType{\fProc{x_i}:\with^{\fPri{\rho_i}} \{ start : \with^{\fPri{\rho'_i}} \{ done : \bullet \} \}}$,
        \item $\fType{\fProc{y_i}:\oplus^{\fPri{\sigma_i}} \{ start : \oplus^{\fPri{\sigma'_i}} \{ done : \bullet \} \}}$
        \item $\fType{\fProc{z_i}:\with^{\fPri{\phi_i}} \{ start : \oplus^{\fPri{\phi'_i}} \{ done : \bullet \} \}}$.
    \end{itemize}
    Applications of Rule~\ruleLabel{typ-res} require that $\fPri{\pi_i = \phi_i}$, $\fPri{\pi'_i = \phi'_i}$, $\fPri{\rho_i = \sigma_i}$, and $\fPri{\rho'_i = \sigma'_i}$, for $1 \leq i \leq 3$.
    Applications of Rules~\ruleLabel{typ-sel} and~\ruleLabel{typ-bra} then require:
    \begin{itemize}
        \item $\fPri{\rho_1 < \rho'_1}$, and $\fPri{\pi_1 < \pi'_1 < \rho'_1,\rho_3}$;
        \item $\fPri{\rho_1 < \pi_2,\rho_2,\pi'_2,\rho'_2}$, $\fPri{\pi_2 < \pi'_2}$, $\fPri{\rho_2 < \rho'_2}$, and $\fPri{\pi'_2 < \rho'_1,\rho'_2}$;
        \item $\fPri{\rho_2 < \pi_3,\rho_3,\pi'_3,\rho'_3}$, $\fPri{\pi_3 < \pi'_3}$, $\fPri{\rho_3 < \rho'_3}$, and $\fPri{\pi'_3 < \rho'_2,\rho'_3}$.
    \end{itemize}
    We verify that these requirements are consistent, so $\fType{\fProc{Sched} \tAPCP \tEmpty}$.
    Hence, $\fProc{Sched} \in \fType{\tAPCP} \setminus \fType{\tACP}$, yet the process is deadlock free (following \cref{t:apcp:df}).
\end{example}

\section{Asynchronous Functional Sessions}
\label{s:lastn}

With \APCP we have established a solid foundation for asynchronous processes that are deadlock free by typing (\Cref{t:apcp:df}). To bring these foundations closer to programming calculi, we consider \LAST: a concurrent \lamcalc-calculus in which asynchronous message-passing is governed using session types~\cite{journal/jfp/GayV10}.
In \LAST, which stands for Linear Asynchronous Session Types, 
channels can form cyclic connections and deadlock freedom is not guaranteed by typing.

The variant \LASTn that we consider here is, in spirit, the functional variant of \AP (\cref{s:ap}):
it is obtained via translation into \AP, resulting in a call-by-name semantics (rather that \LAST's call-by-value semantics) and an explicit treatment of variable substitution.
{Prior works ensure deadlock freedom for synchronous variants of \LAST by extending the type system with priorities~\cite{conf/forte/PadovaniN15,journal/lmcs/KokkeD23}.
Rather than accommodating these contributions to the asynchronous setting, as in \cref{s:apcp}, for \LASTn we develop an alternative approach to deadlock freedom:
}
\LASTn translates into \AP, and operational correctness of the translation guarantees that well typedness under $\fType{\tAPCP}$ implies deadlock freedom for the source program.

We illustrate the call-by-name semantics and explicit substitutions in \LASTn using a simple example.

\begin{example}
    Using standard \lamcalc-calculus notation, the following sequence of term reductions ($\reddM$) and structural congruences ($\scM$) illustrate the call-by-name semantics and explicit substitution ($\fFunc{\tSub{\ldots}}$) of \LASTn:
    \begin{align*}
        \fFunc{\big( \tAbs x . x ~ ( \tAbs y . y ) \big) ~ \big( ( \tAbs w . w ) ~ ( \tAbs z . z ) \big)}
        &\reddM
        \fFunc{\big( x ~ ( \tAbs y . y ) \big) \tSub{ \big( ( \tAbs w . w ) ~ ( \tAbs z . z ) \big)/x }}
        \\
        &\scM
        \fFunc{( x \tSub{ \big( ( \tAbs w . w ) ~ ( \tAbs z . z ) \big)/x } ) ~ ( \tAbs y . y )}
        \\
        &\reddM
        \fFunc{\big( ( \tAbs w . w ) ~ ( \tAbs z . z ) \big) ~ ( \tAbs y . y )}
        \\
        &\reddM
        \fFunc{( w \tSub{ ( \tAbs z . z )/w } ) ~ ( \tAbs y . y )}
        \reddM
        \fFunc{( \tAbs z . z ) ~ ( \tAbs y . y )}
        \\
        &\reddM
        \fFunc{z \tSub{ ( \tAbs y . y )/z }}
        \reddM
        \fFunc{\tAbs y . y}
    \end{align*}
    Notice how every function application immediately evolves into an explicit substitution, instead of first reducing the argument to a value as a call-by-value semantics would.
\end{example}

\LASTn programs are configurations $\fFunc{C},\fFunc{D},\ldots$ of parallel threads executing functional terms $\fFunc{M},\fFunc{N},\ldots$ connected by buffered channels for asynchronous message-passing.
The semantics of configurations is defined in terms of a reduction relation, denoted $\fFunc{C} \reddC \fFunc{D}$.
We write $\reddC*$ for the reflexive, transitive closure of $\reddC$, and $\fFunc{C} \nreddC$ if there is no $\fFunc{D}$ such that $\fFunc{C} \reddC \fFunc{D}$.
Next, we briefly discuss the types of \LASTn, and thereafter focus on the deadlock-freedom guarantee through translation.
We refer to~\cite{journal/lmcs/vdHeuvelP24} for details about the language and type system of \LASTn.

Types can be divided into functional types $\fType{T},\fType{U}$ and session types $\fType{S}$:
\begin{align*}
    \begin{array}[t]{@{}r@{}lrl@{\kern2ex}lrl@{\kern2ex}lrl@{\kern2ex}lrl@{\kern2ex}l@{}}
        \fType{T},\fType{U} ::= {} &
        \fType{T \times U} & \text{pair}
        & \sepr &
        \fType{T \lolli U} & \text{function}
        & \sepr &
        \fType{1} & \text{unit}
        & \sepr &
        \fType{S} & \text{session}
        \\
        \fType{S} ::= {} &
        \fType{\excl T . S} & \text{send}
        & \sepr &
        \fType{\ques T . S} & \text{receive}
        & \sepr &
        \fType{\oplus \{ i : T \}_{i \in I}} & \text{select}
        & \sepr &
        \fType{\with \{ i : T \}_{i \in I}} & \text{branch}
        & \sepr &
        \fType{\tEnd}
    \end{array}
\end{align*}
Configurations may contain at most one main thread with a return type, denoted $\fFunc{\tMain M}$, and arbitrarily many unit-typed child threads, denoted $\fFunc{\tChild N}$.
Configurations are then typed $\fType{\Gamma \tC{\phi} \fFunc{C} : T}$.
The annotation~$\fFunc{\phi}$ can be $\fFunc{\tMain}$ or $\fFunc{\tChild}$, denoting whether $\fFunc{C}$ contains a main thread or not, respectively.
The typing context $\fType{\Gamma}$ is a list of variable-type assignments $\fType{\fFunc{x}:U}$, and $\fType{T}$ is the configuration's return type ($\fType{T} = \fType{1}$ if $\fFunc{\phi} = \fFunc{\tChild}$).

We translate configurations into \AP processes.
Our translation crucially relies on the typing of configurations, in particular to correctly translate buffers.
Note that \AP processes have no functional behavior, so as usual our translation is parametric on a dedicated name on which the source configuration's return type can be observed.
In the following we describe the translation focussing on types, operational correctness, and deadlock freedom;
again, we refer to~\cite{journal/lmcs/vdHeuvelP24} for details on other aspects.

\begin{definition}[Translation]
\label{d:lastntrans}
    The translation of typed \LASTn configurations into \AP processes is denoted $\fProc{\transConfProc{z}{\fType{\Gamma \tC{\phi} \fFunc{C} : T}}}$, where $\fProc{z}$ is the name on which $\fType{T}$ can be observed.
\end{definition}

Although this translation does not return typed processes, it does preserve typing.
Notice that \LASTn and \AP have different types, and so the connection between \LASTn configurations and \AP processes (\Cref{d:lastntrans}) can be captured by a translation on types:  it reflects how the translation on configurations adds additional synchronizations that ensure that the behavior of configurations is soundly captured by the ``more concurrent'' nature of \AP (i.e., the translation does not add behavior not present in the source configuration).
This way, the following definition illustrates the work done by the translation to ensure operational soundness (discussed hereafter).
Given $\fType{\Gamma \tC{\phi} \fFunc{C} : T}$, we refer to types in $\fType{\Gamma}$ as \emph{context types}.

\begin{definition}[Translation: Types]
\label{d:lasttypes}
    The translation of \LASTn types is denoted $\fType{\transType{T}}$, and $\fType{\transTypeDual{T}}$ for context types.
    They are defined mutually, by induction on the structure of $\fType{T}$.
    \begin{align*}
        \fType{\transTypeDual{T}} &\deq \fType{\bullet \tensor \ol{\transType{T}}}
        &
        \\
        \fType{\transType{T \times U}} &\deq \fType{\ol{\transTypeDual{T}} \tensor \ol{\transTypeDual{U}}}
        &
        \fType{\transType{T \lolli U}} &\deq \fType{\transTypeDual{T} \parr \transType{U}}
        &
        \fType{\transType{1}} &\deq \fType{\bullet}
        \\
        \fType{\transType{{!}T.S}} &\deq \fType{\bullet \tensor \transTypeDual{T} \parr \ol{\transTypeDual{S}}}
        &
        \fType{\transType{\oplus \{ i : S_i \}_{i \in I}}} &\deq \fType{\bullet \tensor \with \{ i : \ol{\transTypeDual{S_i}} \}_{i \in I}}
        &
        \fType{\transType{\tEnd}} &\deq \fType{\bullet \tensor \bullet}
        \\
        \fType{\transType{{?}T.S}} &\deq \fType{\ol{\transTypeDual{T}} \tensor \ol{\transTypeDual{S}}}
        &
        \fType{\transType{\with \{ i : S_i \}_{i \in I}}} &\deq \fType{\oplus \{ i : \ol{\transTypeDual{S_i}} \}_{i \in I}}
        &
    \end{align*}
    Context type translation extends naturally to typing contexts $\fType{\transTypeDual{\Gamma}}$.
\end{definition}

\begin{figure}[t]{The translation of $\fType{(T \times S) \lolli {!}T.S}$, given in \cref{x:lastn:transTypes}, in detail.}{f:lastn:transTypes}
    \begin{align*}
        & \fType{( \bullet \tensor {}}
        &&
        \tikzmark{x:lastn:transTypes:1s}
        \tikzmark{x:lastn:transTypes:11s}
        \tikzmark{x:lastn:transTypes:11e}
        &&
        \text{announce pair ready}
        \\
        & \q[1] \fType{( \bullet \tensor {}}
        &&
        \tikzmark{x:lastn:transTypes:12s}
        \tikzmark{x:lastn:transTypes:121s}
        \tikzmark{x:lastn:transTypes:121e}
        &&
        \text{announce substitution ready}
        \\
        & \q[2] \fType{\ol{\transType{T}}}
        &&
        \tikzmark{x:lastn:transTypes:122s}
        \tikzmark{x:lastn:transTypes:122e}
        &&
        \text{first component}
        \\
        & \q[1] \fType{) \parr {}}
        &&
        \tikzmark{x:lastn:transTypes:12e}
        &&
        \text{receive first component}
        \\
        & \q[1] \fType{\bullet \tensor {}}
        &&
        \tikzmark{x:lastn:transTypes:13s}
        \tikzmark{x:lastn:transTypes:13e}
        &&
        \text{announce substitution ready}
        \\
        & \q[1] \fType{\ol{\transType{S}}}
        &&
        \tikzmark{x:lastn:transTypes:14s}
        \tikzmark{x:lastn:transTypes:14e}
        &&
        \text{second component}
        \\
        & \fType{) \parr {}}
        &&
        \tikzmark{x:lastn:transTypes:1e}
        &&
        \text{receive parameter}
        \\
        & \fType{\bullet \tensor {}}
        &&
        \tikzmark{x:lastn:transTypes:2s}
        \tikzmark{x:lastn:transTypes:2e}
        &&
        \text{trigger buffer}
        \\
        & \fType{( \bullet \tensor {}}
        &&
        \tikzmark{x:lastn:transTypes:3s}
        \tikzmark{x:lastn:transTypes:31s}
        \tikzmark{x:lastn:transTypes:31e}
        &&
        \text{announce substitution ready}
        \\
        & \q[1] \fType{\ol{\transType{T}}}
        &&
        \tikzmark{x:lastn:transTypes:32s}
        \tikzmark{x:lastn:transTypes:32e}
        &&
        \text{payload}
        \\
        & \fType{) \parr {}}
        &&
        \tikzmark{x:lastn:transTypes:3e}
        &&
        \text{receive payload provider}
        \\
        & \fType{\bullet \parr {}}
        &&
        \tikzmark{x:lastn:transTypes:4s}
        \tikzmark{x:lastn:transTypes:4e}
        &&
        \text{await substitution ready}
        \\
        & \fType{\transType{S}}
        &&
        \tikzmark{x:lastn:transTypes:5s}
        \tikzmark{x:lastn:transTypes:5e}
        &&
        \text{continuation}
    \end{align*}
    \begin{tikzpicture}[overlay,remember picture]
        \draw ([yshift=2ex]pic cs:x:lastn:transTypes:1s) to [square left brace] (pic cs:x:lastn:transTypes:1e);
        \draw ([yshift=1ex,xshift=.1cm]pic cs:x:lastn:transTypes:1e) to ([yshift=1ex,xshift=.4cm]pic cs:x:lastn:transTypes:1e);
        \draw ([yshift=2ex]pic cs:x:lastn:transTypes:2s) to [square left brace] (pic cs:x:lastn:transTypes:2e);
        \draw ([yshift=1ex,xshift=.1cm]pic cs:x:lastn:transTypes:2e) to ([yshift=1ex,xshift=.4cm]pic cs:x:lastn:transTypes:2e);
        \draw ([yshift=2ex]pic cs:x:lastn:transTypes:3s) to [square left brace] (pic cs:x:lastn:transTypes:3e);
        \draw ([yshift=1ex,xshift=.1cm]pic cs:x:lastn:transTypes:3e) to ([yshift=1ex,xshift=.4cm]pic cs:x:lastn:transTypes:3e);
        \draw ([yshift=2ex]pic cs:x:lastn:transTypes:4s) to [square left brace] (pic cs:x:lastn:transTypes:4e);
        \draw ([yshift=1ex,xshift=.1cm]pic cs:x:lastn:transTypes:4e) to ([yshift=1ex,xshift=.4cm]pic cs:x:lastn:transTypes:4e);
        \draw ([yshift=2ex]pic cs:x:lastn:transTypes:5s) to [square left brace] (pic cs:x:lastn:transTypes:5e);
        \draw ([yshift=1ex,xshift=.1cm]pic cs:x:lastn:transTypes:5e) to ([yshift=1ex,xshift=.4cm]pic cs:x:lastn:transTypes:5e);
        \draw ([yshift=2ex,xshift=.2cm]pic cs:x:lastn:transTypes:11s) to [square left brace] ([xshift=.2cm]pic cs:x:lastn:transTypes:11e);
        \draw ([yshift=1ex,xshift=.3cm]pic cs:x:lastn:transTypes:11e) to ([yshift=1ex,xshift=.6cm]pic cs:x:lastn:transTypes:11e);
        \draw ([yshift=2ex,xshift=.2cm]pic cs:x:lastn:transTypes:12s) to [square left brace] ([xshift=.2cm]pic cs:x:lastn:transTypes:12e);
        \draw ([yshift=1ex,xshift=.3cm]pic cs:x:lastn:transTypes:12e) to ([yshift=1ex,xshift=.6cm]pic cs:x:lastn:transTypes:12e);
        \draw ([yshift=2ex,xshift=.2cm]pic cs:x:lastn:transTypes:13s) to [square left brace] ([xshift=.2cm]pic cs:x:lastn:transTypes:13e);
        \draw ([yshift=1ex,xshift=.3cm]pic cs:x:lastn:transTypes:13e) to ([yshift=1ex,xshift=.6cm]pic cs:x:lastn:transTypes:13e);
        \draw ([yshift=2ex,xshift=.2cm]pic cs:x:lastn:transTypes:14s) to [square left brace] ([xshift=.2cm]pic cs:x:lastn:transTypes:14e);
        \draw ([yshift=1ex,xshift=.3cm]pic cs:x:lastn:transTypes:14e) to ([yshift=1ex,xshift=.6cm]pic cs:x:lastn:transTypes:14e);
        \draw ([yshift=2ex,xshift=.2cm]pic cs:x:lastn:transTypes:31s) to [square left brace] ([xshift=.2cm]pic cs:x:lastn:transTypes:31e);
        \draw ([yshift=1ex,xshift=.3cm]pic cs:x:lastn:transTypes:31e) to ([yshift=1ex,xshift=.6cm]pic cs:x:lastn:transTypes:31e);
        \draw ([yshift=2ex,xshift=.2cm]pic cs:x:lastn:transTypes:32s) to [square left brace] ([xshift=.2cm]pic cs:x:lastn:transTypes:32e);
        \draw ([yshift=1ex,xshift=.3cm]pic cs:x:lastn:transTypes:32e) to ([yshift=1ex,xshift=.6cm]pic cs:x:lastn:transTypes:32e);
        \draw ([yshift=2ex,xshift=.4cm]pic cs:x:lastn:transTypes:121s) to [square left brace] ([xshift=.4cm]pic cs:x:lastn:transTypes:121e);
        \draw ([yshift=1ex,xshift=.5cm]pic cs:x:lastn:transTypes:121e) to ([yshift=1ex,xshift=.8cm]pic cs:x:lastn:transTypes:121e);
        \draw ([yshift=2ex,xshift=.4cm]pic cs:x:lastn:transTypes:122s) to [square left brace] ([xshift=.4cm]pic cs:x:lastn:transTypes:122e);
        \draw ([yshift=1ex,xshift=.5cm]pic cs:x:lastn:transTypes:122e) to ([yshift=1ex,xshift=.8cm]pic cs:x:lastn:transTypes:122e);
    \end{tikzpicture}
\end{figure}

\begin{example}
    \label{x:lastn:transTypes}
    We illustrate the translation of types by means of an example:
    \begin{align*}
        \fType{\transType{(T \times S) \lolli {!}T.S}}
        &= \fType{\transTypeDual{T \times S} \parr \transType{{!}T.S}}
        \\
        &= \fType{( \bullet \tensor \ol{\transType{T \times S}} ) \parr \bullet \tensor \transTypeDual{T} \parr \ol{\transTypeDual{S}}}
        \\
        &= \fType{( \bullet \tensor \ol{\ol{\transTypeDual{T}} \tensor \ol{\transTypeDual{S}}} ) \parr \bullet \tensor ( \bullet \tensor \ol{\transType{T}} ) \parr \ol{\bullet \tensor \ol{\transType{S}}}}
        \\
        &= \fType{( \bullet \tensor \transTypeDual{T} \parr \transTypeDual{S} ) \parr \bullet \tensor ( \bullet \tensor \ol{\transType{T}} ) \parr \bullet \parr \transType{S}}
        \\
        &= \fType{( \bullet \tensor ( \bullet \tensor \ol{\transType{T}} ) \parr \bullet \tensor \ol{\transType{S}} ) \parr \bullet \tensor ( \bullet \tensor \ol{\transType{T}} ) \parr \bullet \parr \transType{S}}
    \end{align*}
    \Cref{f:lastn:transTypes} breaks down the resulting \AP type and explains it in terms of the associated behavior of a process translated from a term implementing the type.
\end{example}

The first property of the translation is type preservation, under $\fType{\tAP}$:

\begin{theorem}[Translation: Type Preservation]
    Suppose given $\fProc{P} = \fProc{\transConfProc{z}{\fType{\Gamma \tC{\phi} \fFunc{C} : T}}}$.
    Then $\fType{\fProc{P} \tAP \transTypeDual{\Gamma} , \fProc{z}:\transType{T}}$.
\end{theorem}

Let us now make precise what we mean by our translation being operationally sound.
The property is based on Gorla's criteria for correct translations~\cite{journal/iandc/Gorla10}.
We focus on operational correctness, which encodes a correspondence between the behavior of a source program and its translation.
Operational soundness then means that any behavior of the translation is reflected by behavior of the source program.
In context of our translation, this means that at any state of execution of a translated configuration, another state can be reached that is the translation of a configuration reachable from the source configuration.
Note that the other direction, where any behavior of the source program is preserved by the translation (operational completeness), holds as well but is omitted for brevity.

\begin{theorem}[Translation: Operational Soundness]
    Suppose given $\fProc{P} = \fProc{\transConfProc{z}{\fType{\Gamma \tC{\phi} \fFunc{C} : T}}}$, and suppose $\fProc{P} \redd* \fProc{Q}$.
    Then there exists $\fFunc{D}$ such that $\fFunc{C} \reddC* \fFunc{D}$ and $\fProc{Q} \redd* \fProc{Q'}$ for $\fProc{Q'} = \fProc{\transConfProc{z}{\fType{\Gamma \tC{\phi} \fFunc{D} : T}}}$.
\end{theorem}

As discussed before, the type system for \LASTn cannot guarantee deadlock freedom by itself.
However, we can rely on \APCP for an indirect result.
Operational soundness is key here, as it guarantees that if a translated configuration is able to reduce, then so is the source configuration.
We are then able to prove deadlock freedom for closed configurations (with empty typing context and unit return type) if their translation is well typed under $\fType{\tAPCP}$.
The idea is that a closed configuration translates into a closed \AP process.
Well typedness under $\fType{\tAPCP}$ (more stringent than under $\fType{\tAP}$) then guarantees deadlock freedom (\cref{t:apcp:df}), which transfers back to the source configuration through operational soundness.
Hence, we state a deadlock-freedom result for \LASTn in the spirit of \cref{t:acp:df,t:apcp:df} (recall: given $\fType{\fProc{P} \tAPCP \tEmpty}$, if $\fProc{P} \nredd$, then $\fProc{P} \sc \fProc{0}$), where the inactive process $\fProc{0}$ corresponds to a main thread of unit value~$\fFunc{\tMain ()}$.
\begin{theorem}[Deadlock Freedom for \LASTn]
\label{t:lastn:df}
    If $\fProc{\transConfProc{z}{\fType{\tEmpty \tCMain \fFunc{C} : 1}}}$ is well typed under $\fType{\tAPCP}$, then $\fFunc{C} \nreddC$ implies $\fFunc{C} \scC \fFunc{\tMain ()}$.
\end{theorem}
Hence, \cref{t:lastn:df} defines a \emph{proof technique} for enforcing deadlock freedom for \LASTn programs, which, given a program, first applies the translation into asynchronous processes and then checks for typability in \APCP (i.e., under~$\fType{\tAPCP}$). Our deadlock-freedom result is conditional in that it is contingent on correct typability in \APCP. A practical procedure for type reconstruction/inference would be needed to correctly associate priorities to the session types obtained using the translation in \cref{d:lasttypes}.

\section{Conclusion}
\label{s:concl}

This paper has presented an overview of recent work on static verification techniques for message-passing processes with asynchronous communication. Based on our results in~\cite{journal/lmcs/vdHeuvelP24}, we have provided a unified presentation for different forms of deadlock en\-force\-ment based on type systems. Our presentation involves four typed languages: \AP, \ACP, \APCP, and \LASTn.

We started by presenting \AP, a simple session $\pi$-calculus with asynchronous communication, whose typing discipline enforces conformance to session protocols but does not exclude deadlocks. As such, \AP is representative of a class of typed process frameworks that make the conscious decision of imposing minimal conditions over the processes/programs that can be typed (cf. \cite{journal/jfp/GayV10,journal/iandc/Vasconcelos12,conf/icfp/ThiemannV16})---in these frameworks, the programmer has wide agency to write typable programs, but also the responsibility of ensuring that their programs enjoy correctness guarantees that go beyond protocol conformance.

Subsequently,  we presented \ACP, an asynchronous variant of Wadler's \CP, which results from \AP via a simple but crucial modification: the two separate typing rules for process composition and restriction in \AP are replaced in \ACP by a single rule that coalesces both constructs following the cut rule in linear logic (Rule~\ruleLabel{typ-cut}). \ACP is based on the logical correspondence that connects session types and \emph{classical} linear logic; its formulation follows the presentation by DeYoung \etal~\cite{conf/csl/DeYoungCPT12}, which was given in the setting of \emph{intuitionistic} linear logic. To our knowledge, an asynchronous version of \CP had not been presented before, so this can be considered an original (yet modest) contribution of this paper. Owing to its logical foundations, \ACP inherits the known expressiveness limitations of its predecessors: it can only enforce deadlock freedom for processes that form tree-like topologies.

\APCP then arises as another variant of \AP, aimed at overcoming these limitations by adopting the key features from priority-based approaches. This ensures that \APCP can enforce deadlock freedom for processes forming cyclic topologies. We briefly discussed how \AP and \APCP provide a suitable foundation for a correct interpretation of a concurrent functional calculus with asynchronous sessions, dubbed \LASTn. The fact that \AP, \ACP, and \APCP determine a wide  spectrum of techniques for  enforcing deadlock freedom is important when considering \LASTn. In fact, \AP suffices for a basic concurrent interpretation of functional sessions; if the interest is in deadlock-free behaviors, then \APCP (and,  to some extent, \ACP) can provide an indirect approach based on a correct typed translation.

Here we have focused on typing disciplines for binary (two-party) protocols.
Further applications of \APCP include the analysis of \emph{multiparty} ($n$-ary) protocols, which involve more than two parties and for which the analysis of deadlock freedom is both important and challenging, especially when processes can interleave actions from different sessions and exchange references to sessions in communications (aka \emph{delegation}), as supported in \AP, \ACP, and \APCP. Interestingly, \APCP provides a basis for the decentralized (static) analysis of process implementations of multiparty protocols, as developed in~\cite{journal/scp/vdHeuvelP22}; this decentralized analysis, in turn, can be adapted also to a setting with dynamic (run-time) verification, to consider (untyped) asynchronous processes whose behavior is governed by a monitoring infrastructure based on session types~\cite{conf/rv/vdHeuvelPD23}. The PhD thesis of the first author provides a unified account of \APCP---its theory and applications---, but also detailed comparisons with related works~\cite{thesis/vdHeuvel24}.

\paragraph{Acknlowedgments }
We are grateful to the organizers and participants of ICE'24 for their comments and to Juan C. Jaramillo for discussions and comments on previous versions of this document.

The research described here has been supported by the Dutch Research Council (NWO) under project No.
016.Vidi.189.046 (`Unifying Correctness for Communicating Software').


\bibliographystyle{eptcs}
\bibliography{refs}

\end{document}